\documentclass[conference]{IEEEtran}
\usepackage{cite}
\usepackage{amsthm}
\usepackage{booktabs}
\usepackage{multirow}
\usepackage{amsmath,amssymb,amsfonts}
\usepackage{algorithm,algorithmic}
\usepackage{graphicx}
\usepackage{textcomp}
\usepackage{calrsfs}

\DeclareMathAlphabet{\pazocal}{OMS}{zplm}{m}{n}
\usepackage{xcolor}
\usepackage{hyperref} 
\usepackage{setspace}
\def\BibTeX{{\rm B\kern-.05em{\sc i\kern-.025em b}\kern-.08em
    T\kern-.1667em\lower.7ex\hbox{E}\kern-.125emX}}

\newlength\myindent
\colorlet{Mycolor2}{green!10!orange!90!gray!90!}
\colorlet{Mycolor}{black}

\usepackage{fancyhdr}
\usepackage{kantlipsum}
\fancypagestyle{plain}{
\fancyhf{}
\fancyhead[C]{Conference on \LaTeX} 

}
\usepackage{eso-pic}

\begin{document}

\AddToShipoutPictureBG*{
\AtPageUpperLeft{
\setlength\unitlength{1in}
\hspace*{\dimexpr0.5\paperwidth\relax}
\makebox(0,-0.75)[c]{\textbf{2020 IEEE/ACM International Conference on Advances in Social Networks Analysis and Mining (ASONAM)}}}}

\title{Hierarchical Overlapping Belief Estimation by Structured Matrix Factorization}

\author{\IEEEauthorblockN{Chaoqi Yang, Jinyang Li, Ruijie Wang, Shuochao Yao,
		Huajie Shao, Dongxin Liu, \\Shengzhong Liu, Tianshi Wang, Tarek F. Abdelzaher}
\textit{University of Illinois at Urbana-Champaign, Urbana, Illinois, 61801}\\
\{chaoqiy2, jinyang7, ruijiew2, syao9, hshao5, dongxin3, sl29, tianshi3, zaher\}@illinois.edu
}

\maketitle

\IEEEoverridecommandlockouts
\IEEEpubid{\parbox{\columnwidth}{\vspace{8pt}
\makebox[\columnwidth][t]{IEEE/ACM ASONAM 2020, December 7-10, 2020}
\makebox[\columnwidth][t]{978-1-7281-1056-1/20/\$31.00~\copyright\space2020 IEEE} \hfill} \hspace{\columnsep}\makebox[\columnwidth]{}}
\IEEEpubidadjcol

\begin{abstract}
	Much work on social media opinion polarization focuses on a {\em flat categorization\/} of stances (or orthogonal beliefs) of different communities from media traces. We extend in this work in two important respects. First, we detect not only points of {\em disagreement\/} between communities, but also points of {\em agreement\/}. In other words, we estimate community beliefs in the presence of {\em overlap\/}. Second, in lieu of flat categorization, we consider {\em hierarchical\/} belief estimation, where communities might be hierarchically divided. For example, two opposing parties might disagree on core issues, but within a party, despite agreement on fundamentals, disagreement might occur on further details. We call the resulting combined problem a {\em hierarchical overlapping belief estimation\/} problem. To solve it, 
	this paper develops a new class of unsupervised Non-negative Matrix Factorization (NMF) algorithms, we call \emph{Belief Structured Matrix Factorization} (BSMF).  
	Our proposed unsupervised algorithm  captures both the latent belief intersections and dissimilarities, as well as hierarchical structure. We discuss properties of the algorithm and evaluate it on both synthetic
	and real-world datasets. In the synthetic dataset, our model reduces error by  $40$\%. In real Twitter traces, it improves accuracy by around $10$\%. The model also achieves $96.08$\% self-consistency in a sanity check.
	
\end{abstract}

\section{Introduction}
\vspace{-1mm}
This paper introduces and solves the novel problem of {\em unsupervised hierarchical overlapping belief estimation\/} that uncovers points of agreement and disagreement among communities (as well as their sub-communities), given their social media posts on a polarizing topic. Most prior work clusters sources or beliefs into flat classes or stances~\cite{kuccuk2020stance}. Instead, we  
focus on scenarios where the underlying social groups disagree on {\em some\/} issues but agree on others (i.e., their beliefs {\em overlap\/}). Moreover, we consider a (shallow) hierarchical structure, where communities can be further subdivided into subsets with their own agreement and disagreement points.

Our work is motivated, in part, by the increasing polarization on social media~\cite{liu2012sentiment}. Individuals tend to connect with like-minded sources~\cite{bessi2016users}, ultimately producing echo-chambers \cite{bessi2016users} and filter bubbles~\cite{bakshy2015exposure}. Tools that could automatically extract social beliefs, and distinguish points of agreement and disagreement among them, might help generate future technologies (e.g., less biased search engines) that summarize information for consumption in a manner that gives individuals more control over (and better visibility into) the degree of bias in the information they consume. 

A key advantage of the solutions described in this paper is that they are {\em unsupervised\/} and mostly {\em language-agnostic\/}. By {\em unsupervised\/}, we mean that our approach does not need prior training, labeling, or remote supervision (in contrast, for example, to deep-learning solutions~\cite{irsoy2014opinion,liu2015fine,wang2017coupled} that usually require labeled data). By (mostly) {\em language-agnostic\/}, we mean that the approach does not use language-specific prior knowledge~\cite{hu2013listening,liu2012sentiment}, distant-supervision~\cite{srivatsa2012mining,weninger2012document}, or prior embedding~\cite{liu2015fine,irsoy2014opinion}. Rather, it relies only on tokenization (the ability to separate individual words). While we test the solution only with English text, we conjecture that the unsupervised nature of the work will facilitate its application to other languages (with the exception of those that do not have spaces between words, such as Chinese and Japanese, because we expect spaces as token separators). An advantage of unsupervised techniques is that they do not need to be retrained for new domains, jargon, or hash-tags (as opposed to techniques that rely, for example, on language embedding or neural networks). To the authors' knowledge, ours is the first unsupervised solution to the problem of hierarchical overlapping belief separation.

The work is a significant generalization of approaches for polarization detection (e.g., \cite{al2017unveiling,bessi2016users,conover2011political,demartini2011analyzing}), that identifies opposing positions in a debate but do not explicitly search for points of agreement. The unsupervised problem addressed in this paper is also different from unsupervised techniques for topic modeling~\cite{ibrahim2018tools,litou2017pythia} and polarity detection~\cite{cheng2017unsupervised,al2017unveiling}. Prior solutions to these problems aim to find {\em orthogonal\/} topic components~\cite{cheng2017unsupervised} or {\em conflicting\/} stances~\cite{conover2011political}. In contrast, we aim to find components that adhere to a given (generic) {\em overlap\/} structure. Moreover, unlike solutions for hierarchical topic decomposition~\cite{weninger2012document,zhang2018taxogen}, we consider not only message content but also user attitudes towards it (e.g., who forwards it), thus allowing for better separation, because posts that share a specific stance are more likely to overlap in the target community (who end up spreading them).


The work is evaluated using both synthetic data as well as real-life data sets, where it is compared to approaches that detect polarity by only considering who posted which claim \cite{al2017unveiling}, approaches that separate messages by content or sentiment analysis \cite{Sentiment140,zhang2014explicit}, approaches that identifies different communities in social hypergraphs \cite{zhou2007learning}, and approaches that detects user stance by density-based feature clustering \cite{darwish2020unsupervised}. The results of this comparison show that our algorithm significantly outperforms the state of the art. An ablation study further illustrates the impact of different design decisions on accomplishing this improvement.
%
%

The rest of the paper is organized as follows. Section~\ref{sec:motivation} formulates the problem and summarizes the solution approach. 
Section~\ref{sec:problemformulation} proposes our new belief structured 
matrix factorization model, and analyzes some model properties.
Section~\ref{sec:experiment} presents an experimental evaluation. We review the related
literature on belief mining and matrix factorization in Section~\ref{sec:related}.
The paper concludes with key observations and a statement on future directions in Section~\ref{sec:conclution}. Note that for the purpose of readability, we have simplified model derivation in the main text, and readers could find technical details in the Appendix.

\section{Problem Formulation}
\vspace{-1mm}
\label{sec:motivation}
Consider an observed data set of posts collected from a social medium, such as Twitter, where each post is associated with a {\em source\/} and with semantic content, called a {\em claim\/}.
Let $\pazocal{S}$ be the set of sources in our data set, and $\pazocal{C}$ be the set of claims made by those sources. While, in this paper, a claim is the content of a tweet (or retweet), the analytical treatment does not depend on this interpretation. 
Let matrix $\mathbf{X}$, of dimension $|\pazocal{S}|\times|\pazocal{C}|$, be a matrix of binary entries denoting who claimed what. If source $S_i$ posted claim $C_j$, then $x_{ij}=1$, otherwise $x_{ij}=0$. 
In general, a claim can be made by multiple sources. 

\subsection{Problem Statement}
\vspace{-0.2mm}
We assume that the set of sources, $\pazocal{S}$, is divided into a small number, $K$, of 
latent social groups, denoted by the subsets $\pazocal{G}_1$, $\pazocal{G}_2$, ..., $\pazocal{G}_K$, that form a tree. In this tree, group $\pazocal{G}_i$ is a child (i.e., a subgroup) of $\pazocal{G}_k$ if $\pazocal{G}_i \subset \pazocal{G}_k$.
Children of the same parent are disjoint groups. Members of group $\pazocal{G}_{k}$ that do not belong to any of its children are denoted by the residual set $\pazocal{G}^-_{k}$.  
Within each group, $\pazocal{G}_k$, $1 \leq k \leq K$, individuals have {\em shared beliefs\/} expressed by a set of claims. A shared belief of a group is a belief espoused by {\em all members\/} of the group. By definition, therefore, a child group inherits the shared beliefs of its parent. The child group may have additional shared beliefs within its group (not shared by the remaining members of the parent). Thus, we define the {\em incremental belief set\/}, $\pazocal{B}_{k}$, of group $\pazocal{G}_{k}$ to be the beliefs held by group $\pazocal{G}_{k}$, beyond what's inherited from its parent. The overall belief set of group $\pazocal{G}_{k}$ is thus the union of incremental beliefs of its ancestors and itself.   
The problem addressed in this paper is to simultaneously (i) uncover all latent groups, $\pazocal{G}_k$, in the tree, and (ii) uncover their incremental belief sets, $\pazocal{B}_k$. 

\begin{figure}[!htb]
    \centering
    \vspace{-2.5mm}
	\includegraphics[width=2.8in]{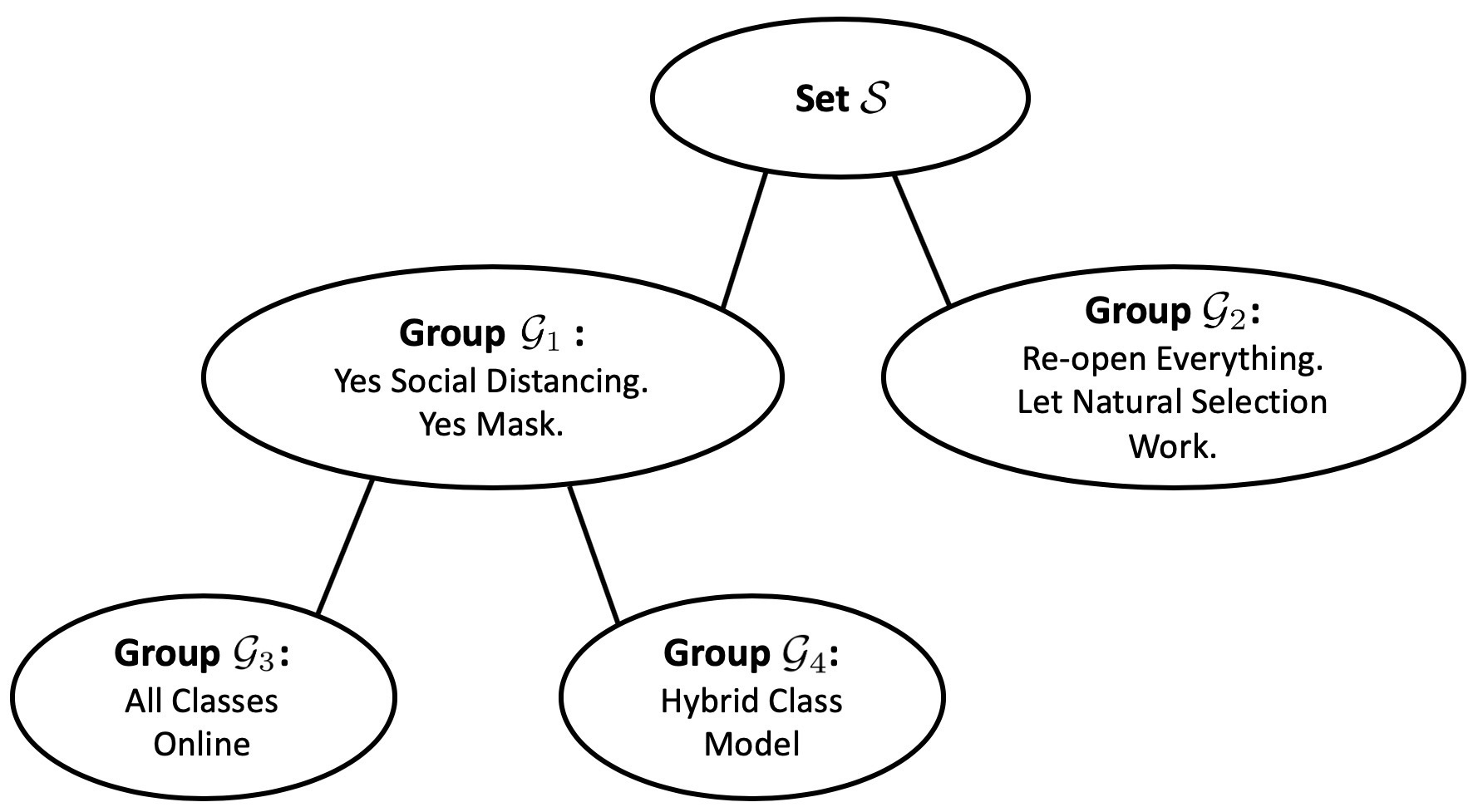}
	\caption{A Notional Example of Hierarchical Overlapping Beliefs.}
	\label{fig:text_demo}
	\vspace{-1.5mm}
\end{figure}

Figure~\ref{fig:text_demo} illustrates an example, inspired by the first wave of the COVID-19 pandemic in 2020. In this figure, a hypothetical community is divided on whether to maintain social distancing ($\pazocal{G}_1$) or reopen everything and let natural selection take place ($\pazocal{G}_2$). Furthermore, while $\pazocal{G}_1$ agree on social distancing, they disagree on some implementation detail, such as whether classes should be entirely online ($\pazocal{G}_3$) or hybrid ($\pazocal{G}_4$).



\vspace{-0.2mm}
\subsection{Solution Approach}
\vspace{-1mm}
We develop a novel non-negative matrix factorization algorithm that decomposes the ``who said what" matrix, $\mathbf{X}$, into (i) a matrix, $\mathbf{U}$, that maps sources to latent groups, (ii) a matrix, $\mathbf{B}$, that maps latent groups to latent incremental belief sets (called the {\em belief structure matrix\/}), and (iii) a matrix, $\mathbf{M}$, that maps latent incremental belief sets to claims. Importantly, since groups and belief sets are {\em latent\/}, the belief structure matrix, $\mathbf{B}$, in essence, specifies the {\em latent structure\/} of the solution space that the algorithm needs to populate with specific sources and claims, thereby guiding factorization. 

\section{Structured Matrix Factorization}
\vspace{-0.5mm}\label{sec:problemformulation}
The novel aspect of our structured matrix factorization is the existence of the structure matrix, $\mathbf{B}$, that represents the relation between the latent groups, $\pazocal{G}_k$ (that we wish to discover), and their incremental belief sets, $\pazocal{B}_k$ (that we wish to discover as well). An element, $b_{ij}$ of the matrix, $\mathbf{B}$, is $1$ if group $\pazocal{G}_i$ adopts the belief set $\pazocal{B}_j$. Otherwise, it is zero. 
In a typical (non-overlapping) clustering or matrix factorization framework, there is an one-to-one correspondence between groups and belief sets, reducing $\mathbf{B}$ to an identity matrix. Structured matrix factorization extends that structure to an arbitrary relation. 
Matrix $\mathbf{B}$ can be thought of as a template relating latent groups (to be discovered) and belief sets (to be identified). It is a way to describe the structure that one wants the factorization algorithm to populate with appropriate group members and claims.
While it might seem confusing to presuppose that one knows the latent structure, $\mathbf{B}$, before the groups and belief sets in question are populated, below we show why this problem formulation is very useful. 

\subsection{An Illustrative Example}
\begin{figure}[htb]
\begin{center}
    \vspace{-0.05in}
	\includegraphics[width=1.2in]{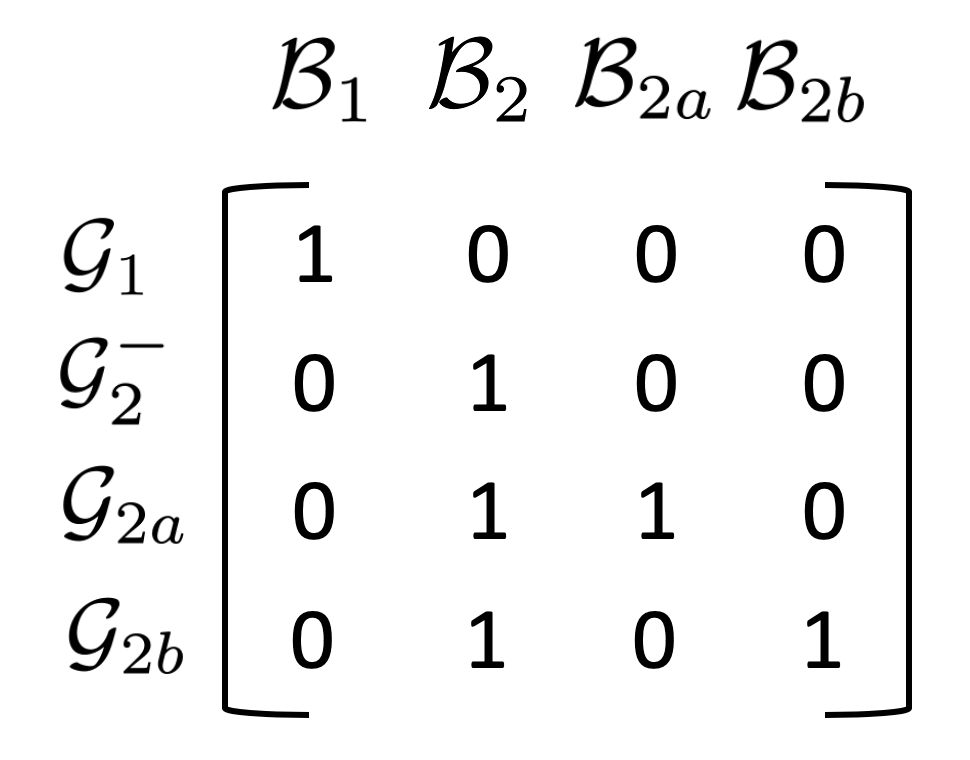}
	\vspace{-0.1in}
	\caption{The Belief Structure Matrix, $\mathbf{B}$.}
	\label{fig:matrix}
	\vspace{-0.2in}
\end{center}
\end{figure}

Consider a conflict involving two opposing groups, say, a minority group $\pazocal{G}_1$ and a majority $\pazocal{G}_2$. Their incremental belief sets are denoted by $\pazocal{B}_1$ and $\pazocal{B}_2$, respectively. The two groups disagree on everything. Thus, sets $\pazocal{B}_1$ and $\pazocal{B}_2$ do not overlap. An agent wants to weaken the majority and conjectures that the majority group might disagree on something internally. Thus, they postulate that group $\pazocal{G}_2$ is predominantly made of subgroups $\pazocal{G}_{2a}$ and $\pazocal{G}_{2b}$. While both subgroups agree on the shared beliefs, $\pazocal{B}_2$, each subgroup has its own incremental belief sets, $\pazocal{B}_{2a}$ and $\pazocal{B}_{2b}$, respectively. 
The structure matrix in Figure~\ref{fig:matrix} represents the belief structure postulated above. 

\noindent
For example, the second column indicates that the belief set $\pazocal{B}_2$ is shared by all members of group 
$\pazocal{G}_2$ (hence, there is a ``1'' in rows of subgroups $\pazocal{G}_{2a}$, $\pazocal{G}_{2b}$, and the residual $\pazocal{G}^-_2$), but that belief sets $\pazocal{B}_{2a}$ and $\pazocal{B}_{2b}$ are unique to subgroups $\pazocal{G}_{2a}$ and $\pazocal{G}_{2b}$, respectively. It is also evident that the beliefs espoused by different groups overlap. For example, from the third and fourth rows, we see that groups $\pazocal{G}_{2a}$ and $\pazocal{G}_{2b}$ overlap in the set of beliefs, $\pazocal{B}_2$. 

An interesting question might be: so, which sources belong to which group/subgroup? What are the incremental belief sets $\pazocal{B}_{2a}$ and $\pazocal{B}_{2b}$ that divide group $\pazocal{G}_2$ (i.e., are shared only by the individual respective subgroups)? What are the shared beliefs $\pazocal{B}_2$ that unite it? What are the beliefs, $\pazocal{B}_1$, of group $\pazocal{G}_1$? These are the questions answered by our structured matrix factorization algorithm whose input is (only) matrix, $\mathbf{X}$, and matrix, $\mathbf{B}$ (Fig.~\ref{fig:matrix}).

\vspace{-0.2mm}
\subsection{Mathematical Formulation} \vspace{-0.8mm}\label{sec:method}
To formulate the hierarchical overlapping belief estimation problem, we introduce the notion of claim {\em endorsement\/}. A source is said to {\em endorse\/} a claim if the source finds the claim agreeable with their belief. Endorsement, in this paper, represents a state of {\em belief\/}, not a physical act. A source might find a claim agreeable with their belief, even if the source did not explicitly post it.
Let the probability that source ${S}_i$ endorses claim ${C}_j$ be denoted by $Pr({S}_i {C}_j)$. We further denote the proposition ${S}_i \in \pazocal{G}_p$ by ${S}_i^p$, and the proposition ${C}_j \in \pazocal{B}_q$ by ${C}_j^q$. Thus,   
$Pr({S}_i^p)$ 
denotes the probability that source ${S}_i \in \pazocal{G}_p$. Similarly, $Pr({C}_j^q)$ 
denotes the probability that claim ${C}_j \in \pazocal{B}_q$. Following the law of total probability:
\begin{align}
Pr({S}_i {C}_j)
&=\mbox{$\sum_{p,q}$} Pr({S}_i {C}_j| {S}_i^p {C}_j^q)Pr({S}_i^p {C}_j^q) \notag\\
&=\mbox{$\sum_{p,q}$} Pr({S}_i {C}_j| {S}_i^p {C}_j^q)Pr({S}_i^p)Pr({C}_j^q).
\label{eq:one}
\end{align}
By definition of the belief structure matrix, $\mathbf{B}$, we say that 
$Pr({S}_i {C}_j| {S}_i^{p} {C}_j^{q}) = 1$
if $b_{pq} = 1$ in the belief structure matrix. Otherwise,  $Pr({S}_i {C}_j| {S}_i^{p} {C}_j^{q}) = 0$.

Let
$u_{ip}= Pr({S}_i^p)$ and $m_{jq}= Pr({C}_j^q)$. Let $u_i$ and $m_j$ be the corresponding vectors, with elements ranging over values of $p$ and $q$ respectively.
Thus, we get:
$Pr({S}_i{C}_j) = 
u_i^\top \mathbf{B} m_j$.
%
%
Let the matrix $\mathbf{X}^G$ be the matrix of probabilities, $Pr({S}_i {C}_j)$, such that element $x^G_{ij} = Pr({S}_i{C}_j)$. 
Thus:
\begin{equation}
\mathbf{X}^G= \mathbf{U} \mathbf{B}\mathbf{M}^\top
\label{eq:b}
\end{equation}
\noindent
where $\mathbf{U}$ is a matrix whose elements are $u_{ip}$ and $\mathbf{M}$ is a matrix whose elements are $m_{jq}$. 
Factorizing $\mathbf{X}^G$, given $\mathbf{B}$, would directly yield $\mathbf{U}$ and $\mathbf{M}$, whose elements are the probabilities we want: elements of matrix $\mathbf{U}$ yield the probabilities that a given source ${S}_i$ belongs to a given group $\pazocal{G}_p$, whereas elements of matrix $\mathbf{M}$ yield the probabilities that a claim ${C}_j$ belongs to a belief $\pazocal{B}_q$. Each source is then assigned to a group and each claim to a belief set, based on the highest probability entry in their row of matrix $\mathbf{U}$ and $\mathbf{M}$, respectively. In practice, $\mathbf{B}$ could be customized up to certain tree depth to meet different granularity of belief estimation.

\subsection{Estimating $\mathbf{X}^G$}
\vspace{-0.5mm}
Unfortunately, we do not really have matrix $\mathbf{X}^G$ to do the above factorization. Instead, we have the {\em observed\/} source-claim matrix $\mathbf{X}$ that is merely a sampling of what the sources endorse. (It is a sampling because a silent source might be in agreement with a claim even if they did not say so.) Using $\mathbf{X}$ directly is suboptimal because it is very sparse. It is desired to estimate that a source endorses a claim even if the source remains silent. We do so on two steps.

\medskip
\noindent
{\em Message interpolation (the M-module):}
First, while source ${S}_i$ might have not posted a specific claim, ${C}_j$, it may have posted similar ones. Thus, even when $x_{ij}=0$, in reality the source may endorse claim ${C}_j$. Let matrix $\mathbf{D}$, of dimension $|\pazocal{C}|\times|\pazocal{C}|$, denote the {\em similarity\/} structure between claims, by some similarity measure. Hence, each element, $d_{ij}$, denotes how similar claims ${C}_i$ and ${C}_j$ are. To keep our approach simple and language-agnostic, as we detail later, we 
use radial basis functions (RBF) that consider only lexical (i.e., bag of words) overlap between the respective claims. We then update $x_{ij}$ if source ${S}_i$ posted claims similar to ${C}_j$, according to matrix, $\mathbf{D}$. This is called {\em message interpolation\/} (the M-module). The exact expression is derived in the Appendix. The output of this module is a matrix $\mathbf{X}^M$. 

\smallskip
\noindent
{\em Social graph convolution (S-module):}
To further improve our estimation of matrix $\mathbf{X}^G$, we assume that sources generally hold similar beliefs to those in their immediate social neighborhood. Thus, we perform a smoothing of matrix 
$\mathbf{X}^M$ by replacing each cell, $x_{ij}$, by a weighted average of itself and the entries pertaining to neighbors of its source, $S_i$, in the social graph.
Let matrix $\mathbf{A}$, of dimension $|\pazocal{S}|\times|\pazocal{S}|$, denote the social graph. Each entry, $a_{ij}$, denotes the influence of user ${S}_i$ on user ${S}_j$.  $\mathbf{A}$ is thus the adjacency matrix of a social network graph. In this
paper, we construct $\mathbf{A}$ by calculating the frequency of each source ${S}_i$ retweeting posts of source ${S}_j$. We call it the {\em retweet graph\/}.
The update of entries of $\mathbf{X}^M$ by smoothing over entries of neighboring sources is called {\em social graph convolution\/} (S-module). The mathematical details are also in the Appendix. It results in an improved estimate, $\mathbf{X}^{MS}$. 

\smallskip
\noindent
We can now take, $\mathbf{X}^G \approx \mathbf{X}^{MS}$, and decompose it as presented in Section~\ref{sec:method}, Equation~(\ref{eq:b}), with $L_1$ and $L_2$ regularizations to enforce sparsity and prevent overfitting. We specify the iterative estimation rule in the Appendix.

\vspace{-0.5mm}
\section{Experiments}
\label{sec:experiment}
In the section, we evaluate \textit{Belief Structured Matrix Factorization (BSMF)} using both a synthetic dataset and real-world Twitter datasets with hierarchical overlapping beliefs. Our model is 
compared to six  
baselines and three model variants.
We elaborate the experimental settings and results below. The code base and additional analytical and experimental results could be found in github \footnote{https://github.com/ycq091044/narrative-detection}.

\vspace{-0.5mm}
\subsection{Synthetic Data} 

\subsubsection{Dataset Construction} In order to understand the behavior of our algorithm in a simplified and controllable setting, we build a synthetic dataset
where two groups of users are created, a minority, $\pazocal{G}_1$ (100 users) of belief set $\pazocal{B}_1$, and a majority, $\pazocal{G}_2$ (300 users) of belief set $\pazocal{B}_2$. The majority includes two subgroups, $\pazocal{G}_{2a}$ and $\pazocal{G}_{2b}$ (100 users each) of incremental belief sets $\pazocal{B}_{2a}$ and $\pazocal{B}_{2b}$, respectively. Essentially, the groups follow the hierarchical structure illustrated in Fig.~\ref{fig:matrix}.
For each of the group, we built disjoint claim corpus, called $c_1$, $c_2$, $c_{2a}$ and $c_{2b}$, respectively. Users were simulated who chose claims to emit based on their corpus or their parent's corpus (we randomly generate 20 claims for each user).
Thus, for example, users in group $\pazocal{G}_{2a}$ could emit claims generated from $c_{2a}$ or from the parent corpus $c_2$, but users in group $\pazocal{G}_1$ only emit claims from corpus $c_1$. In sum, 400 users and 8,000 claims were created.
In this experiment, we
do not impose social relations. Instead, we use the identity matrix for the adjacency $\mathbf{A}$. 

\subsubsection{Method Comparison} 
For this experiment, the factorization
uses the belief structure matrix in Fig.~\ref{fig:matrix}. Two simpler variants are introduced: (i) the first variant substitutes $\mathbf{B}$ with an identity matrix, and takes a standard NMF formulation $\mathbf{X}^G=\mathbf{U}\mathbf{M}^\top$; (ii) the second variant substitutes $\mathbf{B}$ with an learnable matrix $\tilde{\mathbf{B}}$, which takes a standard non-negative matrix tri-factorization (NMTF) form, $\mathbf{X}^G=\mathbf{U}\tilde{\mathbf{B}}\mathbf{M}^\top$.
Obviously, NMTF offers more freedom. However, the need to learn parameters of matrix $\tilde{\mathbf{B}}$ can cause overfitting. 
We use the same regularization settings for NMF, NMTF and our BSMF to make sure the comparison is fair. Empirically, after $150\sim200$ 
iterations, all three methods converge. The predicted belief set label for each claim
is given by the index of the maximum value in this final representation from matrix, $\mathbf{M}$.\footnote{In practice, we permute the labels and pick the best matching as a result, since our approach does clustering not classification.}

\subsubsection{Results of 200 Rounds} 
We run each model for 200 times and BSMF consistently outperforms NMF and NMTF. The
average accuracy for BSMF, NMF and NMTF are $97.34\%$, $93.78\%$, $95.54\%$, respectively. As might be expected, specifying matrix, $\mathbf{B}$, guides subsequent factorization to a better result compared to both NMF and NMTF.

\subsection{Real-world Twitter Trace: \textit{Eurovision2016}}
To illustrate versatility of our structured matrix factorization approach, we start with a simplified example where we break sources into two subgroups of overlapping beliefs. The example allows easy comparison with the prior state of the art on polarization that explicitly addresses this common case.  

\subsubsection{Dataset} We use the Eurovision2016 dataset, borrowed from \cite{al2017unveiling}. Eurovision2016 contains tweets about the Ukrainian singer, Jamala, who won the Eurovision (annual) song contest in 2016. Her success was a
surprise to many as the expected winner had been from Russia according to pre-competition polls. 
The song was on a controversial political topic, about
story of deportation of Crimean Tatars by Soviet Union forces in the 1940s.
Tweets related to Jamala were collected within five days of the contest. Basic statistics are reported in Table~\ref{tb:statistics}. 
As pre-processed in \cite{al2017unveiling}, the most popular 1,000 claims were manually annotated. 
They were separated into 600 pro-Jamala, 239
anti-Jamala, and 161 neutral claims.

In the context of the dataset, the entire set of sources is regarded as a big group, $\pazocal{G}_1$, with belief set, $\pazocal{B}_1$, agreed among all users. Upon the disagreements, the group is further divided into three disjoint groups,  group $\pazocal{G}_{1a}$ (with inherited belief $\pazocal{B}_{1}$ and incremental belief $\pazocal{B}_{1a}$), group $\pazocal{G}_{1b}$ (with inherited belief $\pazocal{B}_{1}$ and incremental belief $\pazocal{B}_{1b}$), and the residual group $\pazocal{G}^-_{1}$ (with belief $\pazocal{B}_{1}$).
In this case, the belief structure matrix is: 
\begin{align}
\mathbf{B}=
\begin{bmatrix}
1&0& 0\\
1&1& 0\\
1&0& 1\\
\end{bmatrix}
\end{align} 

\noindent
where rows correspond to groups $\pazocal{G}^-_1$, $\pazocal{G}_{1a}$, and $\pazocal{G}_{1b}$, respectively, whereas columns correspond to belief sets, $\pazocal{B}_1$, incremental sets, $\pazocal{B}_{1a}$ and $\pazocal{B}_{1b}$, respectively.


\begin{table}[t] \footnotesize
	\centering
	\caption{Basic Statistics for Two Twitter Datasets}
	\begin{tabular}{c|cccc}
		\toprule
		\textbf{Dataset} & \textbf{\# sources} &  \textbf{\# claims} & \textbf{\# all tweets} & \textbf{\# retweets}\\
		\midrule
		Eurovision2016 	&3,514	& 5,812	&	9,868   &   6,001	  \\
		Global Warming 	& 14,752	& 7,030	& 16,418   & 9,341  \\
		\bottomrule
	\end{tabular}
	\label{tb:statistics}
	\vspace{-3mm}
\end{table}

\subsubsection{Baselines} We carefully select seven baseline methods that encompass different perspectives on belief separation:
\begin{itemize}
	\item \emph{Random}: A trivial baseline that annotates posts randomly, giving
	equal probability to each label.
	\item \emph{DBSCAN}\cite{darwish2020unsupervised}: A density-based clustering technique that extracts user-level features and performs DBSCAN for unsupervised user stance detection in Twitter. In this paper, we use DBSCAN and then map the user stance to claim stances with majority voting, as a baseline.
	\item \emph{Sentiment140}\cite{Sentiment140} and SANN \cite{zhang2014explicit}:  Content-aware solutions based on language or sentiment models. In the implementation, each of the claims is a query through Sentiment140 API, which responds with a polarity score. SANN similarly outputs three polarity labels.
	
	\item \emph{H-NCut} \cite{zhou2007learning}: The method views the bipartite structure of 
	the source-claim network as a hypergraph, where claims
	are nodes and sources are hyperedges.
	The problem is thus seen as a 
	hypergraph community detection problem, where community nodes represent posts. We implement \emph{H-NCut}, a  
	hypergraph normalized cut algorithm.
	\item \emph{Polarization} \cite{al2017unveiling}: A baseline that uses an NMF-based solution for social network belief extraction to separate biased and neutral claims.
	\item  \emph{NMTF}: A baseline with a learnable mixture matrix. We compare
	our model 
	with it to demonstrate that pure learning without a prior is not enough to unveil the true belief overlap
	structure in real-world applications.
\end{itemize}


\noindent
Different variants of BSMF are also evaluated to verify the effectiveness of message 
similarity interpolation (the M-module) and social graph convolution (the S-module). BSMF incorporates both modules. 
Models without the M-module or the S-module are
named $\mbox{BSMF}_{M^-}$
and $\mbox{BSMF}_{S^-}$, respectively,  while $\mbox{BSMF}_{MS^-}$ denotes the model without either module.

\subsubsection{Evaluation Metrics}
We evaluate claim separation, since only claim labels are accessible. We use the Python \emph{scikit-learn} package to help with the evaluation. Multiple metrics are employed. 
 {\em Macro-evaluation}
simply calculates the mean of the metrics, giving equal weight to each class. It is 
used to highlight model performance of infrequent classes. {\em Weighted-metrics} account 
for class imbalance by computing the average of metrics in which each class score is 
weighted by its presence in the true data sample. Standard {\em precision, recall and 
f1-score} are considered in both scenarios.
Note that weighted averaging may produce an f1 that is not between precision and recall.

\begin{table}[t] \footnotesize
	
	\caption{Macro- and Weighted- Metrics Comparison (Eurovision2016)}
	\begin{tabular}{l|ccc|ccc}
		\toprule
		& \multicolumn{6} {c} {\bfseries Eurovision 2016} \\
		\cmidrule{2-7}
		\textbf{Models}& \multicolumn{3} {c|} {\bfseries Macro} &\multicolumn{3} {c} {\bfseries Weighted}\\
		\cmidrule{2-7}
		& \textbf{Prec.}            & \textbf{Recall}       & \textbf{F1}  & \textbf{Prec.}            & \textbf{Recall}       & \textbf{F1}\\
		\midrule
		Random     & 0.340         & 0.339        & 0.319 & 0.436       & 0.352        & 0.368\\
		Sentiment140         & 0.425       &   0.377    & 0.339 & 0.413        & 0.384        & 0.354\\
		SANN        & 0.410       & 0.393         & 0.334 & 0.453        & 0.370        & 0.365\\
		H-NCut          & 0.462  & 0.481 &   0.413     & 0.518         & 0.487        & 0.474\\
		DBSCAN & 0.513 & 0.514 & 0.488 & 0.556 & 0.528 & 0.520 \\
		Polarization         & 0.531 & 0.497 & 0.443 & 0.588 & 0.491 & 0.463\\
		NMTF & 0.531 & 0.527 & 0.471 & 0.585 & 0.526 & 0.491\\
		\midrule
		$\mbox{BSMF}_{MS^-}$     & 0.565 & 0.516 & 0.460 & 0.601 & 0.549 & 0.492\\
		$\mbox{BSMF}_{M^-}$     & 0.656 & 0.623 & 0.578 & 0.741 & 0.582 & 0.605\\
		$\mbox{BSMF}_{S^-}$    & 0.586 & 0.592 & 0.574 & 0.664 & \underline{0.660} & \underline{0.645}\\
		BSMF     & \underline{0.678} & \underline{0.643} & \underline{0.596} & \underline{0.768} & 0.602 & 0.623\\
		\bottomrule
	\end{tabular}
	\label{tb:eurovision} \\
	*Prec. stands for precision.
	\vspace{-3mm}
\end{table}

\subsubsection{Result of Eurovision2016} 
The comparison results are shown in Table~\ref{tb:eurovision}. It is not surprising that all 
baselines beat Random. Overall, matrix factorization methods
work well for this problem. With both the \textit{M-module} and \textit{S-module}, our BSMF algorithm
ranks the top in terms of all metrics.
Among other baselines, Sentiment140 and SANN work poorly for this task, because (i) they use
background language models that are pre-trained on another corpus; and (ii) they do not user dependency information, which matters in real-world data. H-NCut and DBSCAN also yield weak performance, sicne they do not consider the underlying overlapping and hierarchical structure. We also notice that 
the NMF-based algorithm actually outperforms NMTF. The reason might be that,
for real-world data, the latent belief structure is harder to capture, and NMTF could be trapped in poor local optima. 

Table~\ref{tb:eurovision2} shows the top 3 
tweets from each belief set ($\pazocal{B}_{1}$, $\pazocal{B}_{1a}$, $\pazocal{B}_{1b}$) estimated by our model. Note that, due to an update of the Twitter API,
the crawled text field is truncated to 140 characters. Our algorithm runs on the text within
that range only. For human readability and interpretability, we manually fill in the rest of the tweet, showing the additional text in \textcolor{Mycolor2}{yellow} (the same for Table~\ref{tb:globalwarming}). Note that, the labels shown in the first column, called {\em Beliefs\/}, are inserted manually after the fact (and not by our algorithm). The algorithm merely does the separation/clustering.

\begin{table}[t]\tiny
	\centering
	\vspace{-1mm}
	\caption{\small Top 3 Tweets From Separated Beliefs (Eurovision2016)}
	\begin{tabular}{p{0.18\columnwidth}|p{0.72\columnwidth}}
		\toprule
		\textbf{(Incremental) Beliefs} & \textbf{Sample Tweets} \\
		\midrule
		\multirow{5}{*}{\textbf{$\pazocal{B}_1$: Agreement}} & \textcolor{Mycolor}{BBC News - Eurovision Song Contest: Ukraine's Jamala wins competition https://t.co/kL8SYOPOYL}\\
		\cmidrule{2-2}
		& \textcolor{Mycolor}{Parents of "\#Ukrainian" Susana \#Jamaludinova - @Jamala
			are \#Russian citizens and prosper in the Russian \#Crimea}\\
		\cmidrule{2-2}
		& \textcolor{Mycolor}{A politically charged ballad by the Ukrainian singer Jamala won the 
@Eurovision
 Song Contest http://nyti.ms/1qlmmNs}\\
		\midrule
		\multirow{5}{*}{\textbf{$\pazocal{B}_{1a}$: Pro-Jamala}} & \textcolor{Mycolor}{@jamala congratulations! FORZA UKRAINE!}\\
		\cmidrule{2-2}
		& \textcolor{Mycolor}{@DKAMBinUkraine: Congratulations @jamala and \#Ukraine!!! You deserved all the 12 points from \#Denmark and the victory, \#workingforDK}\\
		\cmidrule{2-2}
		& \textcolor{Mycolor}{@NickyByrne: Well done to Ukraine and @jamala} \\
		\midrule
		\multirow{5}{*}{\textbf{$\pazocal{B}_{1b}: $Anti-Jamala}} &  \textcolor{Mycolor}{jamala The song was political and agaisnt The song contest rules shows NATO had influence on jury decision.}\\
		\cmidrule{2-2}
		& \textcolor{Mycolor}{@VictoriaLIVE @BBCNews @jamala Before voting we rated it worst song in the contest. Not changed my mind.}\\
		\cmidrule{2-2}
		& \textcolor{Mycolor}{\textcolor{Mycolor}{@JohnDelacour So @jamala has violated TWO ESC rules - the song is not new, and it includes political content.} \textcolor{Mycolor2}{Result MUST be annulled}}\\
		\bottomrule
	\end{tabular}
	\label{tb:eurovision2}
	\vspace{-2mm}
\end{table}

\subsection{A More Complicated Scenario: Global Warming} 

We consider a more complicated scenario in this section, with a majority group, $\pazocal{G}_1$, and a minority group, $\pazocal{G}_2$, of beliefs $\pazocal{B}_1$ and $\pazocal{B}_2$ that do not overlap. Since more data is expected on $\pazocal{G}_1$ (by definition of majority), we opt to further divide it into subgroups $\pazocal{G}_{1a}$ and $\pazocal{G}_{1b}$, who (besides believing in $\pazocal{B}_1$) hold the incremental belief sets $\pazocal{B}_{1a}$ and $\pazocal{B}_{1b}$, respectively.
The corresponding belief structure is reflected by the belief matrix:
\begin{align} \small
\mathbf{B}=\begin{bmatrix}
1&0&0& 0\\
1&1&0& 0\\
1&0&1& 0\\
0&0&0& 1\\
\end{bmatrix}
\end{align} 
\noindent
where rows represent groups 
$\pazocal{G}^-_1$, 
$\pazocal{G}_{1a}$, 
$\pazocal{G}_{1b}$,
and $\pazocal{G}_{2}$
and columns represent the belief sets 
$\pazocal{B}_1$,
$\pazocal{B}_{1a}$,
$\pazocal{B}_{1b}$,
and $\pazocal{B}_2$.
The matrix is not an identity matrix because beliefs overlap (e.g, groups $\pazocal{G}^-_1$, 
$\pazocal{G}_{1a}$, and
$\pazocal{G}_{1b}$ share belief  $\pazocal{B}_1$). It also features a hierarchical subdivision of $\pazocal{G}_1$, into $\pazocal{G}^-_1$, 
$\pazocal{G}_{1a}$, and
$\pazocal{G}_{1b}$.

\subsubsection{Datasets} We apply this belief structure to an unlabeled dataset, Global Warming, which is crawled in real time
with the \emph{Apollo Social Sensing Toolkit\footnote{http://apollo2.cs.illinois.edu/}}. This dataset is about a twitter discussion of global warming in the wake of Australia wildfires that ravaged the continent, in September 2019, where at least 17.9 million acres of forest have burned in the fire. Our goal is to identify and separate posts according to the above abstract belief structure. 

\begin{table}[t]\tiny
	\centering
	\caption{\small Top 3 Tweets From Separated Beliefs (Global Warming)}
	\begin{tabular}{p{0.20\columnwidth}|p{0.70\columnwidth}}
		\toprule
		\textbf{(Incremental) Beliefs} & \textbf{Sample Tweets} \\
		\midrule
		& Australia's top scientists urge government to do more on global warming https://t.co/NclFqGKXE1\\
		\cmidrule{2-2}
		{\textbf{$\pazocal{B}_1$: Global Warming / urge response}} & Australia's most prestigious scientific organisation has added to growing pressure on Prime Minister Scott Morrison over climate \textcolor{Mycolor2}{change policy, calling on the government to "take stronger action" in response to the bushfire crisis}\\
		\cmidrule{2-2}
		& "Have we now reached the point where at last our response to global warming will be driven by engineering and economics rather than ideology and idiocy?" \#auspol\\
		\midrule
		& As long as the ALP keep accepting ‘donations’ (bribes) from the climate change deniers the fossil fuel industry, who spent \textcolor{Mycolor2}{millions and millions  spreading lies about global warming, they have zero creditibility when they talk about phasing out fossil fuel}\\
		\cmidrule{2-2}
		{\textbf{$\pazocal{B}_{1a}$: Global Warming / fossil fuel}} & To mitigate the effects of climate change, we must do away with fossil fuel burning as they are the major contributors of \textcolor{Mycolor2}{global warming}. \\
		\cmidrule{2-2}
		& Turnbull: “The world must, and I believe will, stop burning coal if we are to avoid the worst consequences of global warming. \textcolor{Mycolor2}{And the sooner the better.” Malcom Turnbull, The Guardian 12 January \#ScottyfromMarketing}\\
		\midrule
		& That time when The Australian misrepresented 
		@JohnChurchOcean
		to say sea level rise wasn’t linked to global warming. After I \textcolor{Mycolor2}{wrote about it, they pulled the story.}\\
		\cmidrule{2-2}
		{\textbf{$\pazocal{B}_{1b}$: Global Warming / sea level}} & Brave global warming researchers are studying sea level rise in the Maldives this morning. https://t.co/aqGtgXAj2t \\
		\cmidrule{2-2}
		& CO2 is a magical gas which causes Lake Michigan water levels to both rise and fall https://t.co/8FrC1Cx2Rm\\
		\midrule
		& CLIMATE’S FATAL FLAW : ‘Greenhouse Gases Simply Do Not Absorb Enough Heat To Cause Global Warming’ – \textcolor{Mycolor2}{“New data and improved understanding now show that there is a fatal flaw in greenhouse-warming theory.”}\\
		\cmidrule{2-2}
		{\textbf{$\pazocal{B}_2$: No Global Warming}} & “Three new research studies confirm that geothermal heat flow, not man-made global warming, is the dominant cause of West \textcolor{Mycolor2}{Antarctic Ice Sheet (WAIS) melting,” writes geologist James Edward Kamis.}\\
		\cmidrule{2-2}
		& Left Media talks about.. Climate Change, Global Warming, But... Jihad is reason for recent Forest Fires in Australia !\\
		\bottomrule
	\end{tabular}
	\label{tb:globalwarming}
	\vspace{-1mm}
\end{table}

Table~\ref{tb:globalwarming} shows the algorithm's assignment of claims to belief groups (only the top 3 claims are shown for space limitations). The first column shows the abstract belief categories 
$\pazocal{B}_1$,
$\pazocal{B}_{1a}$,
$\pazocal{B}_{1b}$,
and $\pazocal{B}_2$.
While the algorithm allocates posts to categories based on the structure of matrix $\mathbf{B}$, for readability, we manually inspect posts assigned to each category in the matrix, and give that category a human-readable name, also shown in the first column. 
For each belief category, the table also shows the top ranked statements.  
The table reveals that sources in our data set are polarized between a group, $\pazocal{G}_1$, that believes in global warming (offering statements that urge a serious response) and a group, $\pazocal{G}_2$, that does not (offering statements that oppose the thesis of man-made global warming). Within group, $\pazocal{G}_1$ (apart from $\pazocal{G}^-_{1}$), there are two subgroups, $\pazocal{G}_{1a}$ and $\pazocal{G}_{1b}$. The former blames the fossil fuel industry, whereas the latter is concerned with rising sea levels. 
While we do not claim to have reached conclusions on global warming, the table shows how structured matrix factorization can fit data sets automatically to useful belief structures, thereby offering visibility into what individuals are concerned with, what actions they agree on, and what they disagree about.

\subsubsection{Quantitative Measurements} Next, we do a sanity check by measuring user grouping consistency. Specifically, we first identify the belief sets (by claim separation) and then assign belief labels to users by having a user inherit the assigned belief set label for each claim they made. The inherited labels are inconsistent if they belong to {\em different\/} groups according to matrix, $\mathbf{B}$. 
For example, if the same user has been assigned belief labels $\pazocal{B}_1$ and $\pazocal{B}_{1a}$, then the labeling is coherent because both represent beliefs of $\pazocal{G}_1$ (remember that a group inherits the beliefs of its parent). If another user is labeled with both $\pazocal{B}_1$ and $\pazocal{B}_2$, then it is apparently wrong, since belief sets $\pazocal{B}_{1}$ and $\pazocal{B}_{2}$ belong to different groups. 

The percentage of coherently labeled users was $96.08$\%. Note that, we do not conduct comparison in this dataset, since most baselines do not uncover hierarchical group/belief structures, whereas those that do generally break up the hierarchy differently (e.g., by hierarchical topic, not hierarchical stance) thus not offering an apples to apples comparison. In future work, we shall explore more comparison options.

\section{Related Work} 
\label{sec:related}
The problem of belief mining has been a subject of study for decades \cite{liu2012sentiment}. Solutions include such diverse approaches as detecting social polarization \cite{al2017unveiling,conover2011political}, opinion extraction \cite{irsoy2014opinion,liu2015fine,srivatsa2012mining}, stance detection \cite{darwish2020unsupervised} and sentiment analysis \cite{hu2013unsupervised,hu2013listening}, to name a few.

Pioneers, like Leman \emph{at el.} \cite{akoglu2014quantifying} and Bishan \emph{at el.} \cite{yang2012extracting}, had used Bayesian models and other basic classifiers to separate social beliefs. On the linguistic side, many efforts
extracted user opinions based on domain-specific phrase chunks \cite{wu2018hybrid}, and temporal expressions \cite{schulz2015small}.
With the help
of pre-trained embedding, like Glove \cite{liu2015fine} or word2vec \cite{wang2017coupled}, deep neural networks (e.g., variants of RNN \cite{irsoy2014opinion, liu2015fine}) emerged as powerful tools (usually with attention modules \cite{wang2017coupled}) for understanding the polarity or sentiment of user messages. In contrast to the above supervised or language-specific solutions, we consider the challenge of developing an \emph{unsupervised} approach.

In the domain of unsupervised algorithms, our problem is different from the related problems of unsupervised topic detection \cite{ibrahim2018tools,litou2017pythia}, sentiment analysis \cite{hu2013unsupervised,hu2013listening}, and unsupervised community detection~\cite{fortunato2016community}. Topic modeling assigns posts to polarities or topic mixtures~\cite{han2007frequent}, independently of actions of users on this content. Hence, they often miss content nuances or context that helps better interpret the stance of the source. 
Community detection~\cite{yang2013overlapping}, on the other hand, groups nodes by their general interactions, maximizing intra-class links while minimizing inter-class links \cite{yang2013overlapping,fortunato2016community}, or partitioning (hyper)graphs \cite{zhou2007learning}. While different communities may adopt different beliefs, this formulation fails to distinguish regions of belief overlap from regions of disagreement.

The above suggests that belief mining must consider both sources (and forwarding patterns) and content. Prior solutions used a source-claim bipartite graph, and determined disjoint polarities by iterative factorization \cite{al2017unveiling,akoglu2014quantifying}. Our work is novel by postulating a more generic and realistic view: social beliefs could overlap and can be hierarchically structured. In this context, we developed a new matrix factorization scheme that considers (i) the source-claim graph \cite{al2017unveiling}; (ii) message word similarity \cite{weninger2012document} and (iii) user social dependency \cite{zhang2013maximizing} in a new class of non-negative matrix factorization techniques to solve the hierarchical overlapping belief estimation problem.

%

The work also contributes to non-negative matrix factorization.
NMF was first introduced
by Paatero and Tapper \cite{paatero1994positive} as the concept of positive matrix
factorization and was popularized by the work of Lee and
Seung \cite{lee2001algorithms}, who gave an interesting interpretation based on parts-based representation.
Since then, NMF has been widely used in various applications, such as pattern recognition \cite{cichocki2009nonnegative}, 
signal processing \cite{buciu2008non}.

Two main issues of NMF have been intensively discussed during the development
of its theoretical properties: solution uniqueness \cite{donoho2004does,klingenberg2009non} and decomposition sparsity 
\cite{moussaoui2005non,laurberg2008theorems}.  By only considering the standard formula $\mathbf{X\approx UM}^\top$,  it is 
usually not difficult to
find a non-negative and non-singular matrix $\mathbf{V}$, such that $\mathbf{UV}$ and 
$\mathbf{V^{-1}M^\top}$ could also be a valid solution.
Uniqueness will be achieved if  $\mathbf{U}$ and $\mathbf{M}$ are
sufficiently sparse or if additional constraints are included \cite{wang2012nonnegative}.  Special constraints have 
been proposed in \cite{mohammadiha2009nonnegative,hoyer2004non} to improve the sparseness of the final representation.  

Non-negative matrix tri-factorization (NMTF) is an extension of conventional NMF (i.e., 
$\mathbf{X\approx UBM^\top}$ \cite{yoo2010orthogonal}). Unconstrained NMTF is theoretically identical to 
unconstrained  NMF. However, when constrained,
NMTF possesses more degrees of freedom \cite{wang2012nonnegative}. NMF on a manifold emerges when the
data lies in a nonlinear low-dimensional submanifold \cite{cai2008non}. Manifold Regularized Discriminative NMF \cite{guan2011manifold,ana2011manifold} were
proposed with special constraints to preserve local invariance, so as to reflect the 
multilateral characteristics.

In this work, instead of including constraints to impose structural properties, 
we adopt a novel belief structured matrix factorization by introducing the 
mixture matrix $\mathbf{B}$. The structure of $\mathbf{B}$ can well
reflect the latent belief structure and thus narrows the search space to a good enough region.

\section{Conclusion} \label{sec:conclution}
In this paper, we proposed a new class of NMF, where the structure of parts is already known (or assumed to follow some generic form). 
Specifically, we introduced a belief structure matrix $\mathbf{B}$, and 
proposed a novel \emph{Belief Structured Matrix Factorization} algorithm, called BSMF, to separate overlapping, hierarchically structured beliefs from large volumes of user-generated messages.  The factorization could be 
briefly formulated as $\mathbf{X}^{MS}\approx \mathbf{UBM}^\top$, where $\mathbf{B}$ is known. 
The model is tested on a synthetic dataset. Further evaluations are conducted on real-world Twitter events. The results show that our algorithm consistently outperform baselines by a great margin. 
We believe this paper could seed a research direction on automatically separating data sets according to arbitrary belief structures to enable more in-depth understanding of social groups, attitudes, and narratives on social media.

\linespread{0.93}
\section*{Acknowledgement}
Research reported in this paper was sponsored in part by DARPA award W911NF-17-C-0099, DTRA award HDTRA118-1-0026, the Army Research Laboratory under Cooperative Agreement W911NF-17-20196. The views and conclusions contained in this document are those of the author(s) and should not be interpreted as representing the official policies of the CCDC Army Research Laboratory, DARPA, DTRA, or the US government. The US government is authorized to reproduce and distribute reprints for government purposes notwithstanding any copyright notation hereon.

\linespread{0.9}
\bibliographystyle{ieee}
\bibliography{references}

\section*{Appendix}
\linespread{1}
\vspace{-1mm}
\subsection{Message Interpolation (M-module)} \label{sec:interpolation}
\vspace{-1mm}
The approximation, $\mathbf{X}^M\approx \mathbf{X}^G$, is developed as follows. 
First, if a source $\pazocal{S}_i$ posted, retweeted, or liked claim $\pazocal{C}_j$ in our data set (i.e., $x_{ij}=1$ in matrix $\mathbf{X}$), then we know that the source endorses that claim (i.e., $x^M_{ij} = 1$ in matrix $\mathbf{X}^M$). The question is, what to do when $x_{ij}=0$? In other words, we need to estimate the likelihood that the source endorses a claim, when no explicit observations of such endorsement were made. We do so by considering the claim similarity matrix $\mathbf{D}$. If source $\pazocal{S}_i$ was observed to endorse claims $\pazocal{C}_k$ similar to $\pazocal{C}_j$, then it will likely endorse $\pazocal{C}_j$ with a probability that depends on the degree of similarity between $\pazocal{C}_j$ and $\pazocal{C}_k$. 
Thus, when $x_{ij}=0$, we can estimate $x^M_{ij}$ by  weighted sum interpolation:
\begin{equation}\label{eq:general}
x^M_{ij} = \sum_{k:~x_{ik}\neq 0} d_{kj}\cdot x_{ik}
\end{equation}

\noindent
To compute matrix $\mathbf{D}$, in this work, we first compute a bag-of-words (BOW) vector $w_j$ for each claim $j$. We then normalize it using vector $L_2$ norm, $\bar{w_j} = w_j/\|w_j\|_2$. 
We select non-zero entries $x_{ij}$ in each row $i$ of $\mathbf{X}$ as \emph{medoids} $\{\bar{w_j}\mid x_{ij}\neq0\}$.
We assume that claims close to any of the \emph{medoids} could also be endorsed by $\pazocal{S}_i$ as well.
Based on that, we use: 
\begin{equation} \label{eq:rbf}
d_{kj}=\phi(\|\bar{w_j}-\bar{w_k}\|)
\end{equation}
in Equation~(\ref{eq:general}).
A Gaussian radial basis function (RBF) is used for $\phi(r)=e^{-(\epsilon r)^2}$. If the resulting 
value of $x^M_{ij}$ is less than 0.2, we regard that it is far from all of the \emph{medoids} and set it back to 0. In the experiment, $\epsilon$ is set 0.5 for sythetic datset and 0.05 for both Eurovision 2016 and Global Warming.

\vspace{-1mm}
\subsection{Social Graph Convolution (S-module)} \label{sec:graphconv}
\vspace{-1mm}
To further improve our estimation of matrix $\mathbf{X}^G$, denoted by $\mathbf{X}^{MS}$, we consider the social dependency matrix $\mathbf{A}$.

The fundamental insight we would like to leverage is that users that are close in the social graph, $\mathbf{A}$, are likely to endorse the same claims, even if an explicit endorsement was not observed in the data set. Thus, we consider the
social dependency matrix $\mathbf{A}$ (user-user retweet frequency) and compute the a degree
matrix $\mathbf{F}$ by summing each row of $\mathbf{A}$ and the random walk 
normalized adjacency is denoted as $\tilde{\mathbf{A}_{rw}} = \mathbf{F}^{-1}\mathbf{A}$. We define our propagation 
operator based on $\tilde{\mathbf{A}_{rw}}$ with self-loop re-normalization, 
$\bar{\mathbf{A}_{rw}} \leftarrow \frac12\tilde{\mathbf{F}_{rw}}^{-1}(\tilde{\mathbf{A}_{rw}} +I)$. Thus, the new source-claim network is given by,
\begin{equation}
\mathbf{X}^{MS} = \bar{\mathbf{A}_{rw}}X^M,
\end{equation}
where each row of $\bar{\mathbf{A}_{rw}}$ adds up to 1. The effect of the propagation operator is to convolve
the information from 1-hop neighbors, while preserving half of the information from itself. Note that,
we deem dependency beyond 1-hop too weak to capture, so we do not consider $\mathbf{A}^n$, where $n>1$. From a 
macroscopic perspective, this social graph convolution recovers some of the possible source-claim 
connections and also enforces the smoothness of matrix $\mathbf{X}^{MS}$. 

\linespread{0.87}
\subsection{Overall Loss and Optimization}
Given a belief mixture matrix, $\mathbf{B}$, we now factorize $\mathbf{X}^{MS}$ to estimate matrices
$\mathbf{U}$ and $\mathbf{M}$ that decide the belief regions associated with sources and claims, respectively. (e.g., the estimated belief for claim $\pazocal{C}_j$ is given by the index of maximum entry in the 
$j_{th}$ row of $\mathbf{M}$). 

\emph{Regularization.} To avoid model overfitting, we include widely used $L_2$ regularization. Also,
we enforce the sparsity of $\mathbf{U}$ and $\mathbf{M}$ by introducing $L_1$ norm.
The overall objective function becomes (defined by the Forbenious-norm),
\begin{align}
J&= \|\mathbf{X}^{MS}-\mathbf{UBM}^\top\|_F^2 + \lambda_1\|\mathbf{U}\|_F^2+\lambda_1\|\mathbf{M}\|_F^2 \notag\\
&+\lambda_2\|\mathbf{U}\|_1+\lambda_2\|\mathbf{M}\|_1.
\end{align}
We rewrite $J$ using matrix trace function $tr(\cdot)$,
\begin{align} \label{eq:loss}
J &= tr(\mathbf{X}^{MS^\top} \mathbf{X}^{MS}) - 2tr(\mathbf{X}^{MS^\top} \mathbf{UBM}^\top) \notag\\
& +tr(\mathbf{UBM}^\top \mathbf{MB}^\top \mathbf{U}^\top) + \lambda_1 tr(\mathbf{U}^\top \mathbf{U}) +\lambda_1tr(\mathbf{M}^\top \mathbf{M})\notag\\
& +\lambda_2\|\mathbf{U}\|_1+\lambda_2\|\mathbf{M}\|_1.
\end{align}

We minimize $J$ by gradient descent. Since only the non-negative region is of our interests, 
derivatives of
$L_1$ norm is differentiable in this setting. By referring to gradient of traces of product with 
constant matrix $A$, $\nabla_\mathbf{X}tr(\mathbf{AX})=\mathbf{A}^\top$ and $\nabla_\mathbf{X}tr(\mathbf{XAX}^\top)=\mathbf{X(A+A}^\top)$, 
the partial derivative of $J$ w.r.t.  $\mathbf{U}$ and $\mathbf{M}$ are calculated as,
\begin{align}
&\nabla_\mathbf{U}= -2\mathbf{X}^{MS}\mathbf{MB}^\top +2\mathbf{UBM}^\top \mathbf{MB}^\top + 2\lambda_1 \mathbf{U} + \lambda_2 \mathbf{1}, \notag\\
&\nabla_\mathbf{M}=-2\mathbf{X}^{MS^\top} \mathbf{UB} +2\mathbf{MB}^\top \mathbf{U}^\top \mathbf{UB} + 2\lambda_1 \mathbf{M}+ \lambda_2 \mathbf{1}. \notag
\end{align}

The gradient matrix $\nabla_\mathbf{U}$ is of dimension $|\pazocal{S}|\times K$, and $\nabla_\mathbf{M}$ is 
of dimension $|\pazocal{C}|\times K$. Estimation step begins by updating $\mathbf{U}\leftarrow \mathbf{U}-\eta\nabla_\mathbf{U}$ 
and $\mathbf{M}\leftarrow \mathbf{M}-\eta\nabla_\mathbf{M}$, and $\eta$ is the step size. Negative values might appear in the
learning process, which are physically meaningless in this problem. Thus, we impose the non-negative constraints
for $\mathbf{U}$ and $\mathbf{M}$ during the update. A \emph{ReLU}-like strategy is utilized: when any entry of
$\mathbf{U}$ or $\mathbf{M}$ becomes negative, it is set to be $\xi$. In the experiment, we set $\xi=10^{-8},~\lambda_1=\lambda_2=10^{-3}$. Note that the initial entry values of 
$\mathbf{U}$ and $\mathbf{M}$ are randomized uniformly from $(0,1)$.

\subsection{Complexity} 
After \textit{M-module} and \textit{S-module}, non-zero entries in the estimated matrix, $\mathbf{X}^{MS}\approx\mathbf{X}^G$, are still far fewer than $|\pazocal{S}|\times |\pazocal{C}|$. 
We consider 
to use sparse matrix multiplications and avoid dense intermediate matrices, which makes the
computation efficient. Note that, $K$ (number of beliefs) is picked according to the dataset, and it typically
satisfies $K \ll min(|\pazocal{S}|,|\pazocal{C}|)$. During the estimation, we generalize 
standard NMF multiplicative update rules \cite{lee2001algorithms} for our
tri-factorization,
\begin{equation}
\eta_{\mathbf{U}} = \frac12\frac{\mathbf{U}}{\mathbf{UBM}^\top \mathbf{MB}^\top},~~\eta_{\mathbf{M}} = \frac12\frac{\mathbf{M}}{\mathbf{MB}^\top \mathbf{U}^\top \mathbf{UB}}.
\end{equation} 

Algorithmically, updating $\mathbf{U}$ and $\mathbf{M}$ takes $O\left(K|\pazocal{S}||\pazocal{C}|\right)$ per iteration. We could also take the
advantages of the structure of $\mathbf{B}$, and reduce the complexity to $O\left(|\pazocal{S}||\pazocal{C}|\right)$, 
identical to typical NMF. The number 
of iterations before the empirical convergence is usually no more than 200 for random initialization,
and thus we claim that our model is scalable and efficient.

\end{document}